\documentclass[useAMS,usenatbib]{mn2e}
\usepackage{graphicx}
\usepackage{color}

\title[Probing the anisotropic local universe with SNe\,Ia]{Probing
  the anisotropic local universe and beyond with SNe\,Ia data}
\author[J. Colin, R. Mohayaee, S. Sarkar \& A. Shafieloo]
 {Jacques Colin$^\dagger$,
  Roya Mohayaee$^\dagger$, 
  Subir Sarkar$^\star$ and
  Arman Shafieloo$^\star$$^\ddagger$\\
  $^\dagger$ UPMC, CNRS, Institut d'Astrophysique de Paris, 
            98 bis Bd. Arago, Paris 75014, France\\
  $^\star$ Rudolf Peierls Centre for Theoretical Physics, 
          University of Oxford, 1 Keble Road, Oxford, OX1 3NP, UK\\
 $^\ddagger$ Institute for the Early Universe, Ewha Womans University, 
          Seoul, 120-750, South Korea}
\date{\today}
\begin{document}

\maketitle

\label{firstpage}

\begin{abstract}
  The question of the transition to global isotropy from our
  anisotropic local universe is studied using the Union~2 catalogue of
  Type~Ia supernovae (SNe\,Ia).  We construct a `residual' statistic
  sensitive to systematic shifts in their brightness in different
  directions and use this to search in different redshift slices for a
  preferred direction on the sky in which the SNe\,Ia are brighter or
  fainter relative to the standard $\Lambda$CDM cosmology. At low
  redshift ($z<0.05$) we find that an isotropic model such as
  $\Lambda$CDM is barely consistent with the SNe\,Ia data at 2--3
  $\sigma$. A maximum likelihood analysis of peculiar velocities
  confirms this finding --- there is a bulk flow of 260~km\,s$^{-1}$
  extending out to $z\sim0.06$, which disagrees with $\Lambda$CDM at
  1--2 $\sigma$. Since the Shapley concentration is believed to be
  largely responsible for this bulk flow, we make a detailed study of
  the infall region: the SNe\,Ia falling away from the Local Group
  towards Shapley are indeed significantly dimmer than those falling
  towards us on to Shapley. Convergence to the CMB rest frame must
  occur well beyond Shapley ($z>0.06$) so this low redshift bulk flow will
  systematically bias any reconstruction of the expansion history of
  the universe. At higher redshifts $z>0.15$ the agreement between the
  SNe\,Ia data and the $\Lambda$CDM model does improve, however, the
  sparseness and low quality of the data means that the latter cannot
  be singled out as the preferred cosmological model.
\end{abstract}

%\pacs{98.80.Es,98.65.Dx,98.62.Sb}

\section{Introduction}
\label{sec:introduction}

Modern cosmology is founded on the cosmological principle which
assumes that the universe is homogeneous and isotropic. The local
universe is however observed to be anisotropic and inhomogeneous,
exhibiting the `cosmic web' of voids and superclusters. This
presumably causes the Local Group of galaxies moves towards a
preferred direction $\ell=276^\circ\pm3, b=30^\circ\pm2$ at
$627\pm22$~km\,s$^{-1}$, as inferred from the dipole anisotropy of the
cosmic microwave background (CMB) radiation \citep{Kogut:1993ag}.

On the other hand the overall isotropy of the CMB (barring the dipole
anisotropy due to our local motion) provides strong evidence for an
isotropic universe on very large scales.\footnote{The WMAP
  observations of anomalies in large-angle anisotropies in the CMB,
  e.g. the hemispherical asymmetry \citep{Eriksen:2003db} and the
  unexpected quadrupole-octupole alignment \citep{de
    OliveiraCosta:2003pu}, have led many to question whether the CMB
  is indeed statistically isotropic
  (e.g. \citet{Copi:2006tu}). However others have argued that these
  anomalies may simply be due to the manner in which the galactic
  foreground was masked \citep{Pontzen:2010yw}. We look to forthcoming
  observations by Planck to resolve this contentious issue.} Where
does the transition between these two regimes occur? While
high-quality data exist at low redshifts and the CMB provides reliable
information at very high redshifts, the data is rather sparse and
mainly of poor quality on the intermediate scales of interest. Here
the SNe\,Ia Hubble diagram is the key source of information and so we
use the comprehensive Union~2 catalogue~\citep{Amanullah:2010vv} to
investigate these important questions.

SNe\,Ia data has been examined previously to test the isotropy of the
universe
\citep{Kolatt:2000yg,Bonvin:2006en,Gordon07,Schwarz07,Gupta08,Koivisto:2007bp,Koivisto:2008ig,Blomqvist:2008ud,Cooray:2008qn,Gupta10,Koivisto:2010dr}
and to determine whether the universe accelerates differently in
different directions. Recently a marginal ($1\sigma$) detection of
anisotropy has been reported on spatial scales where dark energy
becomes important \citep{Cooke:2009ws,Antoniou:2010gw}. Clearly,
better quality and more complete surveys are needed before any firm
conclusions can be drawn. However, although these detections are not
significant by themselves, a puzzling and perhaps accidental feature
is the alignment of the detected anisotropy with the the CMB dipole
direction. In this work, we demonstrate that the alignment at low
redshift is due to the attraction of huge structures such as the
Shapley supercluster. At high redshift, the alignment seems to become
statistically insignificant but given the sparse and poor quality
data, no strong conclusions can be drawn.

On small scales, the CMB dipole and the bulk flow are aligned, which
is unsurprising as the common source of both these motions is very
likely the anisotropic distribution of matter in the local universe.
However, a bulk flow much larger than expected has been found
extending out to at least 120 Mpc
\citep{Watkins:2008hf,Lavaux:2008th},
\footnote{\Citet{Kashlinsky:2008ut,Kashlinsky:2009dw} have reported an
  even faster coherent flow out to at least $\sim 800$~Mpc, using
  measurements of the kinematic Sunyaev-Zeldovich effect in galaxy
  clusters.} which is a $\sim2-3\sigma$ fluctuation in a $\Lambda$CDM
model since convergence to the CMB rest frame ought to occur at much
smaller scales in this model. At low redshifts, we study the bulk flow
using two different methods: first by a method of `smoothing and
residuals' (Sections~\ref{sec:meth} and \ref{sec:res}) and secondly by
a maximum likelihood analysis (Section~\ref{sec:likelihood}). We show
that these two methods are in good agreement at small redshifts ($z <
0.05$) and confirm that there is indeed a discrepancy between the
$\Lambda$CDM model prediction for the bulk flow and the SNe\,Ia
observations.

The Shapley concentration at $z \simeq 0.035-0.055$ is believed to be
the main source of our large bulk motion. We study the infall region
around Shapley (Section \ref{sec:shapley}) and demonstrate that
SNe\,Ia beyond Shapley are systematically brighter than expected in an
isotropic universe (as they are falling towards Shapley), while
SNe\,Ia between the Local Group and Shapley are systematically dimmer
(as they are also falling towards Shapley, but away from us). This
result is obtained using our smoothing and residuals scheme.

At high redshift, the $\chi^2$ statistic cannot be used since the bulk
flow becomes small relative to the Hubble expansion rate. Using the
method of smoothing and residuals for $z > 0.05$ we find that an
isotropic model such as $\Lambda$CDM is consistent with the
data. However, the poor quality of the data means that anisotropic
models cannot be eliminated.

Since the high redshift results may be contaminated by the low
redshift data, we perform a `cosmic tomography' --- the data is sliced
up in redshift and the question of isotropy is studied separately for
each slice. The differential analysis is then complemented by a
commulative analysis to establish the correlation between different
shells. We present our conclusions in Section \ref{sec:conclusion}.

\section{The Union 2 supernovae catalogue}

The Union 2 catalogue~\citep{Amanullah:2010vv} contains 557 SNe\,Ia of
which 165 are at $z < 0.1$. This is the largest compilation of SNe\,Ia
to date from several different surveys including
Union~\citep{Kowalski:2008ez}, CfA~\citep{Hicken:2009df} and
SDSS~\citep{Kessler:2009ys}, in addition to data on 6 new SNe\,Ia
reported by \citet{Amanullah:2010vv}. While efforts have been made to
obtain good control over the systematics, it is recognised that
combining data from different surveys can introduce additional
biases. Nevertheless this is the best data set available at present.

\begin{figure}
\center{
\includegraphics[width=0.5\textwidth]{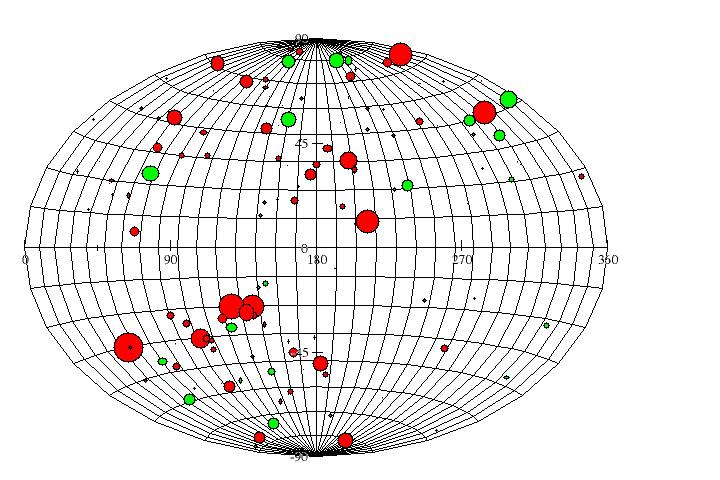}
\includegraphics[width=0.5\textwidth]{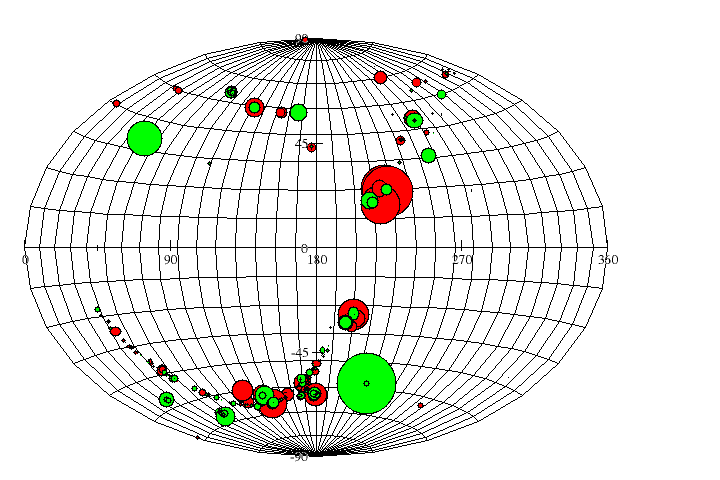}
\includegraphics[width=0.5\textwidth]{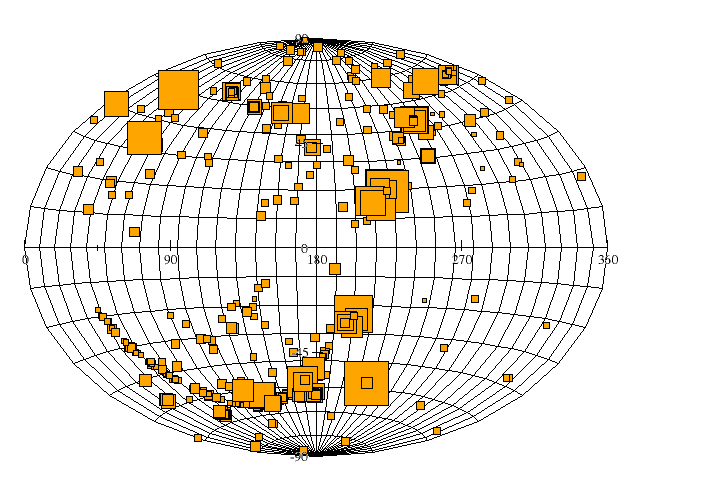}}
\caption{Top panel: Aitoff-Hammer presentation of the Union~2 data for
  $z<0.06$. The red spots represent the data points with distance
  moduli, $\mu_\mathrm{data}$, larger than the values,
  $\mu_\mathrm{{\Lambda}CDM}$, predicted by $\Lambda$CDM and the green
  spots are those with $\mu_\mathrm{data}<\mu_\mathrm{{\Lambda}CDM}$;
  the spot size is a relative measure of the discrepancy. A dipole
  anisotropy is visible around the direction $b=-30^\circ,\ell=96^\circ$ (red
  points) and its opposite direction $b=30^\circ,\ell=276^\circ$ (small green
  points), which is the direction of the CMB dipole.  The middle panel
  is the same plot for $z>0.06$. The data seems to be homogeneously
  distributed at small redshifts but suffers from a significant
  selection effect at large redshifts. The bottom panel shows the full
  Union 2 data set ($0.015<z<1.5$) but now the size of the spots
  correspond to the observational uncertainties --- a clear
  correlation is seen with the model-data discrepancy.}
\label{fig:union2}
\end{figure}

In Figure~\ref{fig:union2}, we illustrate by the size of the spots the
discrepancy between the luminosity distance of Union~2 SNe\,Ia and the
value predicted by the standard $\Lambda$CDM model, for $z < 0.06$
(top panel) and $z > 0.06$ (middle panel). The red spots indicate
SNe\,Ia further away than the $\Lambda$CDM prediction, while the green
spots indicate those which are closer. In the bottom panel of
Figure~\ref{fig:union2} the size of the spots represents the
observational uncertainties. We note that there is a clear correlation
between the latter and the discrepancy between the luminosity
distances and the $\Lambda$CDM model.

%-------------
%-----------

\begin{figure}
\includegraphics[width=0.46\textwidth]{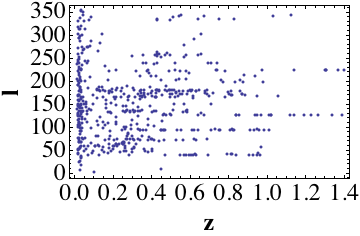}
\includegraphics[width=0.46\textwidth]{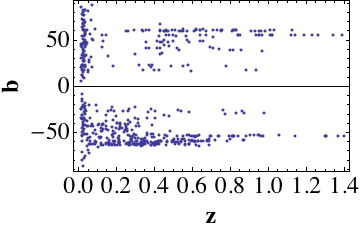}
\caption{The distribution in galactic longitude ($\ell$) and latitude
  ($b$) of the Union 2 SNe\,Ia, as a function of the redshift $z$. At
  high redshift, there are significant selection effects and it is
  difficult to draw any definite conclusions about the anisotropy of
  the universe; however at small redshift ($z < 0.06$) the data is
  quite homogeneous and robust tests of local anisotropy can be made.}
\label{fig:union2-lb}
\end{figure}

As seen in Figure~\ref{fig:union2-lb}, the data is quite homogeneously
distributed over the sky at redshifts $z < 0.06$;  
however at higher redshifts the data becomes increasingly sparse.

%-------------
%-----------

\section{The method of residuals}
\label{sec:meth}

If the assumption of the isotropy of the universe is correct then any
residuals remaining from the subtraction of the isotropic model from the data
should be distributed randomly around zero (assuming the systematics
are under control). In this work, we fit the standard $\Lambda$CDM
model to the Union 2 data \citep{Amanullah:2010vv}, subtract the
best-fit model from the data and then build a statistic to analyse the
residuals. This analysis involves the following steps:

First, we consider a $H (z)$ parameterization (viz. $\Lambda$CDM) and
a set of SNe\,Ia data $\mu_i (z_i, \theta_i, \phi_i)$ --- we choose to
use galactic coordinates ($\ell$, $b$) for $\theta_i, \phi_i$. At low
redshifts $z \ll 1$, the Hubble law implies a linear relationship
between distance and redshift so the choice of cosmological model is
irrelevant, however this becomes important at high redshift.

Second, we obtain the best fit form of $\tilde H (z)$ and the
corresponding best-fit distance moduli $\tilde\mu(z)$, then
construct the error-normalised difference of the data from the model,
\begin{equation}
  q_i (z_i, \theta_i, \phi_i) = \frac{\mu_i(z_i, \theta_i, \phi_i) 
    - \tilde\mu_i(z_i, \theta_i, \phi_i)}{\sigma_i(z_i, \theta_i, \phi_i)}\;, 
\end{equation}
as in previous work
\citep{Perivolaropoulos:2008yc,Shafieloo:2010xm}. Here, $\mu_i(z_i,
\theta_i, \phi_i)$ and $\sigma_i (z_i, \theta_i, \phi_i)$ represent
the distance modulus and the relative error bar of the $i$'th data
point, and $\tilde\mu_i(z_i, \theta_i, \phi_i)$ is the expected
distance modulus in the assumed theoretical model at $z_i$.
Henceforth we work with these residuals, $q_i (z_i, \theta_i,
\phi_i)$, and consider their angular distribution on the sky.

Third, we define a measure $Q (\theta, \phi)$ on the surface of a
sphere of unit radius using these residuals:
\begin{equation}
  Q (\theta,\phi) = \sum_{i=1}^N q_i (z_i, \theta_i, \phi_i)
   W (\theta, \phi, \theta_i, \phi_i)\;, 
\end{equation} 
where $N$ is the number of SNe\,Ia data points and $W (\theta, \phi,
\theta_i,\phi_i)$ is a weight function (or window function) in our 2D
smoothing over the surface of the sphere. We define this weight by the
Gaussian distribution:
\begin{equation}
  W (\theta, \phi, \theta_i, \phi_i) = \frac{1}{\sqrt{2\pi}\delta} 
   \exp\left[-\frac{L(\theta, \phi, \theta_i, \phi_i)^2}{2\delta^2}\right]\;,
\end{equation} 
where $\delta$ is the width of smoothing and $L (\theta, \phi,
\theta_i, \phi_i)$ is the distance on the surface of a sphere of unit
radius between two points with spherical coordinates $(\theta, \phi)$
and $(\theta_i, \phi_i)$:
\begin{equation}
  L (\theta, \phi, \theta_i, \phi_i) = 2 \arcsin \frac{R}{2}\;, 
\end{equation} 
where 
\begin{eqnarray*}
R=\Big(\big[\sin(\theta_i)\cos(\phi_i)-\sin(\theta)\cos(\phi) \big]^2 +\\
\big[\sin(\theta_i)\sin(\phi_i)-\sin(\theta)\sin(\phi) \big]^2 
+ \big[\cos(\theta_i)-\cos(\theta)\big]^2 \Big) ^{1/2}.
\end{eqnarray*} 
Thus, any anisotropy in the data in any specific direction will
translate into $Q (\theta, \phi)$ being significantly less or more
than zero, depending on the quality of the data.

Finally we adopt a value for $\delta$, calculate $Q (\theta, \phi)$ on the
whole surface of the sphere and find the minimum and maximum of this
quantity, $Q (\theta_\mathrm{min}, \phi_\mathrm{min})$ and $Q
(\theta_\mathrm{max}, \phi_\mathrm{max})$. Our statistical measure of
anisotropy is based on the difference
\begin{equation} 
  \Delta Q_\mathrm{data} = Q (\theta_\mathrm{max}, \phi_\mathrm{max}) 
   - Q (\theta_\mathrm{min}, \phi_\mathrm{min}) ,
\label{DeltaQ}
\end{equation}
i.e. a large value of $\Delta Q_\mathrm{data}$ implies significant
anisotropy. To estimate the significance, we simulate 1000
realizations of the data with the same angular positions on the sky
and the same error bars (assuming a Gaussian distribution), determine
$\Delta\,Q_j$ and then do a simple test to determine the significance
of our results for the real data. Any anisotropy in the SNe\,Ia sample
would be more significant if $\Delta\,Q_\mathrm{data}$ is {\em larger}
than most of $\Delta\,Q_j$s for the simulated data. From the resulting
frequency distribution of this statistic one can derive a P-value,
defined as the probability that, given the null hypothesis, the value
of the statistic is larger than the one observed. We remark that in
defining this statistic one has to be cautious about {\it a
  posteriori} interpretations of the data; a particular feature
observed in the real data may be very unlikely, but the probability of
observing {\em some} feature may be quite large--- see
\citet{Spergel:2003cb}, \citet{Lewis:2003qm} and the discussion in
\citet{Hamann:2009bz}.

We would like to point out here some additional advantages of our
method. First, the only parameter in this analysis is the value of
$\delta$ --- the width of gaussian smoothing on the surface of the 2D
sphere. Its value cannot affect the results when testing the
consistency of the data with an isotropic universe, since we are
dealing only with the residuals and the real data is compared with its
simulated samples in exactly the same manner. Second, the isotropy of
the data can be checked in a completely {\em model-independent} manner
and the significance of the findings can also be checked to avoid any
misinterpretation of the data. Third, the method is able to detect
anisotropy even in small patches of the sky if there is sufficient
data available. In fact our method is not limited to finding only
dipoles in the sky but can be useful if there is a neighboring local
void or any unexpected bulk flow up to modest redshifts in a specific
direction. Moreover, because of working with the normalised residuals,
the method is insensitive to those regions of the sky where there is
no data, which makes it relatively bias-free.  Hence it is
particularly suitable for $z > 0.05$ where there are rather few data
points. Although in this analysis we focus on the isotropic model of
the universe, our method can be used to test any other model, in
particular anisotropic models.

We wish to emphasise the difference between testing the consistency of
a cosmological model with the data, and finding the model that best
describes the universe. Depending on the quality, quantity and
distribution of the data there may be several (degenerate) models that
are all consistent with the data. For example, we may find that the
isotropic model of universe fits the data, but that the data is so
sparse at high redshifts that in many patches of the sky we do not
have even a single data point, so an anisotropic model which has
features in these directions can also fit the data.

\section{Test of isotropy using the method of residuals}
\label{sec:res}

We now use the method of smoothing and residuals to test the isotropy
of the universe using SNe\,Ia data at different redshifts and for
different values of the smoothing parameter $\delta$. We divide the
Union 2 data~\citep{Amanullah:2010vv} into 5 bins at $z < 0.1$ and 7
bins at $z>0.1$ so as to have sufficient number of data points in each
bin --- see Table~\ref{table:data}. In each case the real data is
compared with the results from 1000 Monte Carlo realisations assuming
an isotropic universe. The underlying cosmological model is assumed to
be a flat $\Lambda$CDM model with $\Omega_\mathrm{0m}=0.27$ (which is
very close to the best-fit flat $\Lambda$CDM model for Union 2 data).

%--------------
\begin{table*}
  \caption{Number of SNe\,Ia per redshift bin in
    the Union 2 catalogue}
\begin{tabular}{ccccccccc}
\hline
\hline
$\Delta\,z$ & 0.015--0.025 & 0.025--0.035 & 0.035--0.045 & 0.045--0.06 & 0.06--0.1 & 0.1--0.2\\
\hline
$N$ & 61 &48 & 18 & 15 & 23 & 55\\
\hline
\hline
$\Delta z$ & 0.2--0.3 & 0.3--0.4 & 0.4--0.5 & 0.5--0.6 & 0.6--0.8 & 0.8--1.4\\
\hline
$N$ & 62 & 63 & 58 & 43 & 51 & 60\\
\hline
\end{tabular}
\label{table:data}
\end{table*}
%-----------

\begin{figure*}
\includegraphics[width=0.49\textwidth]{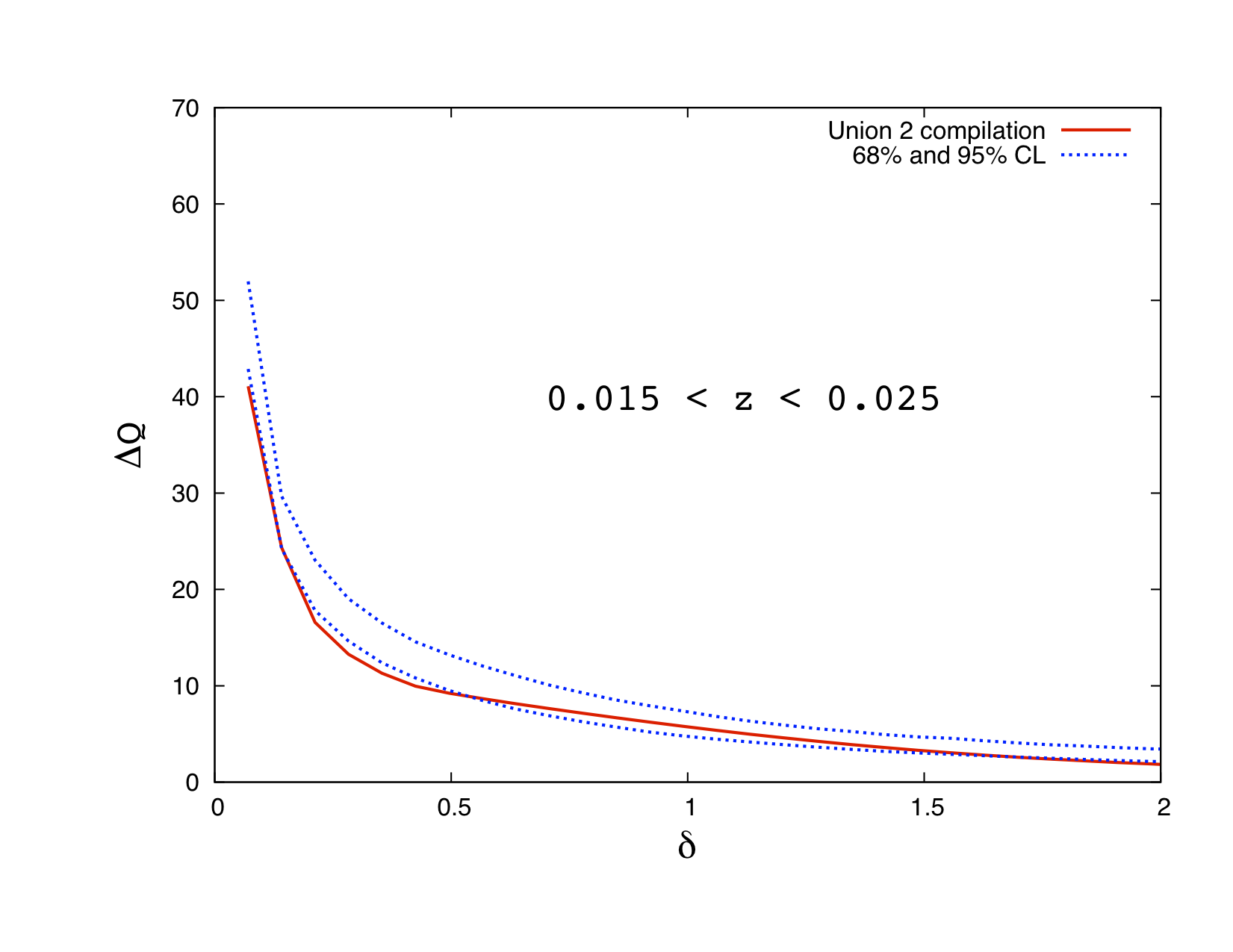}
\includegraphics[width=0.49\textwidth]{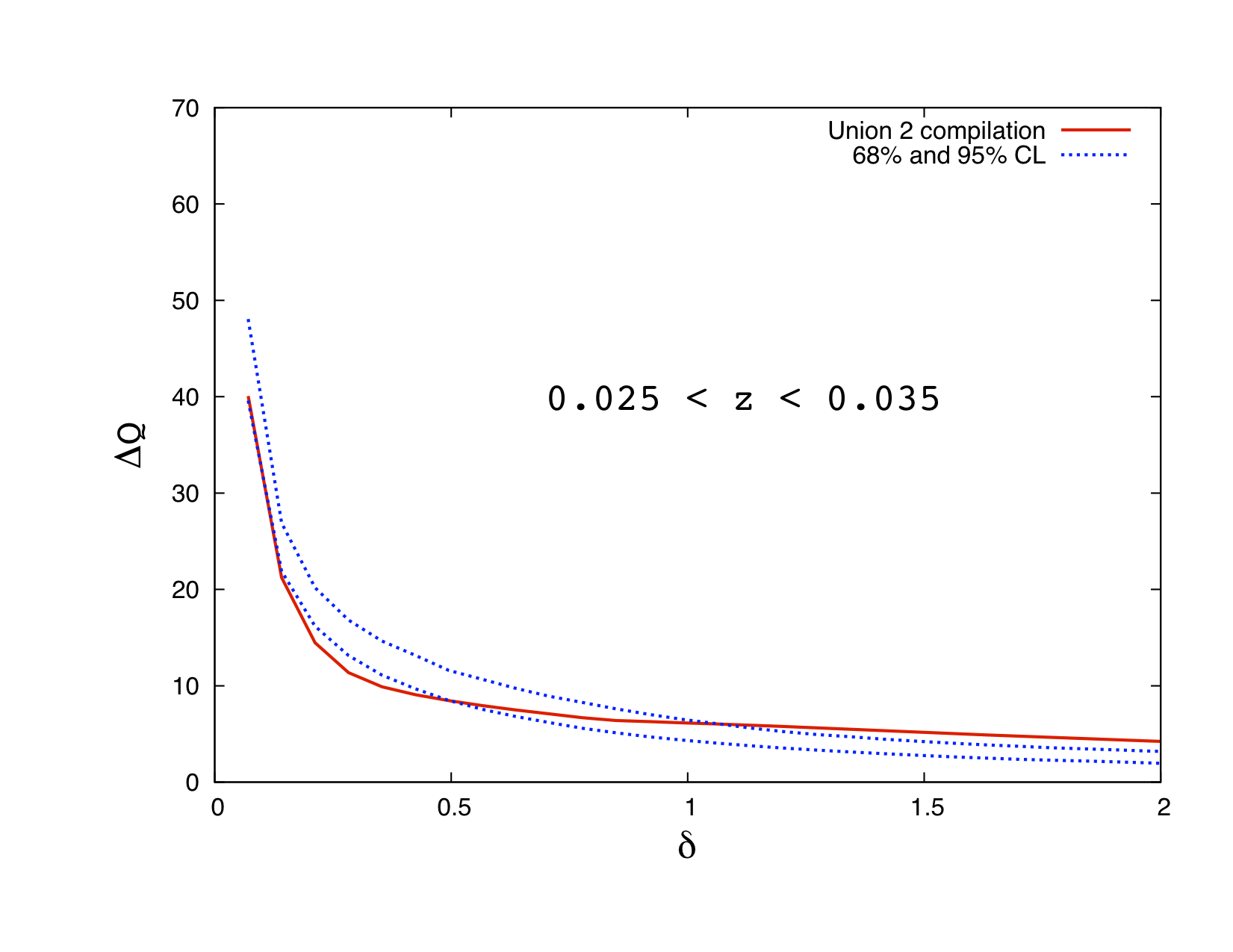}
\includegraphics[width=0.49\textwidth]{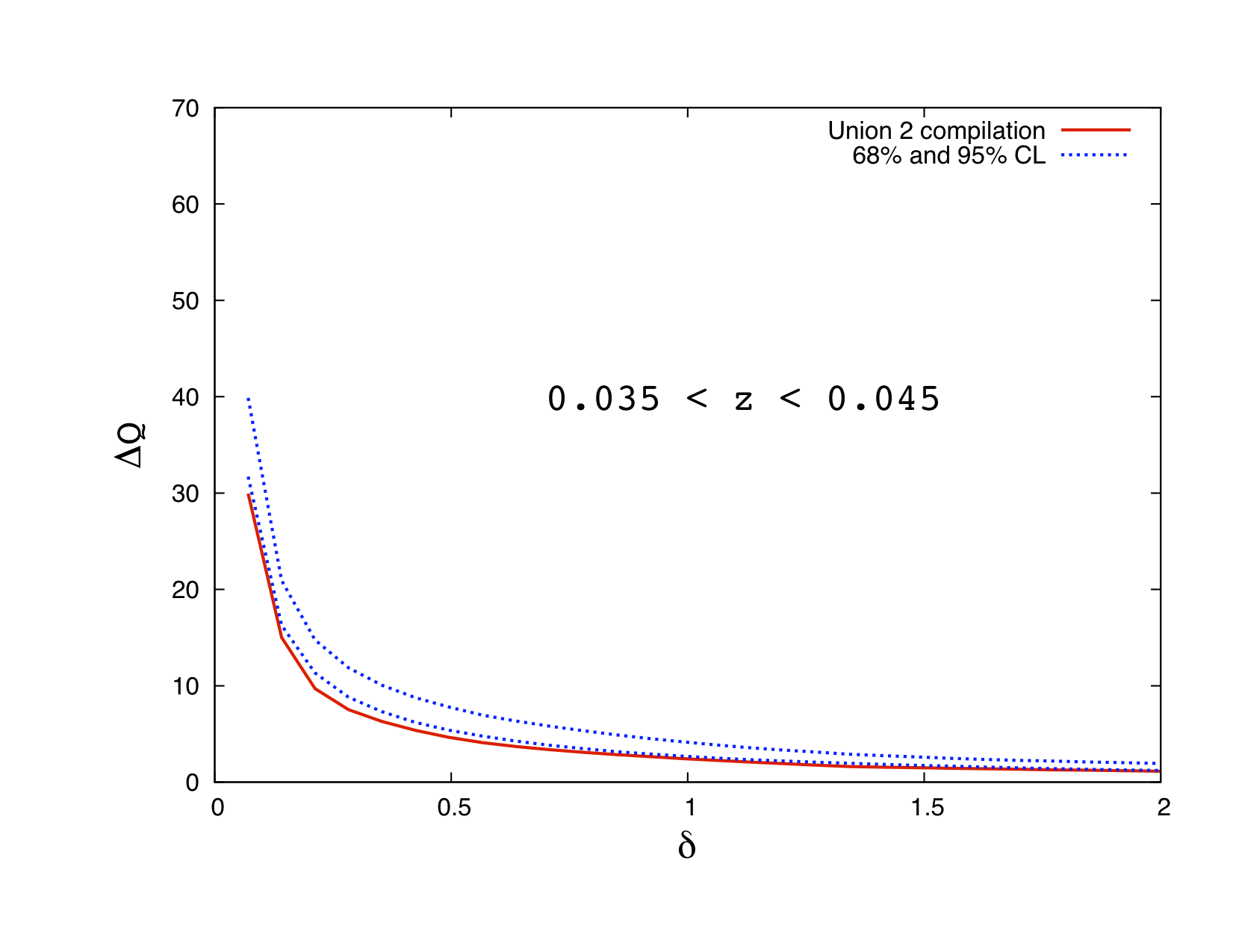}
\includegraphics[width=0.49\textwidth]{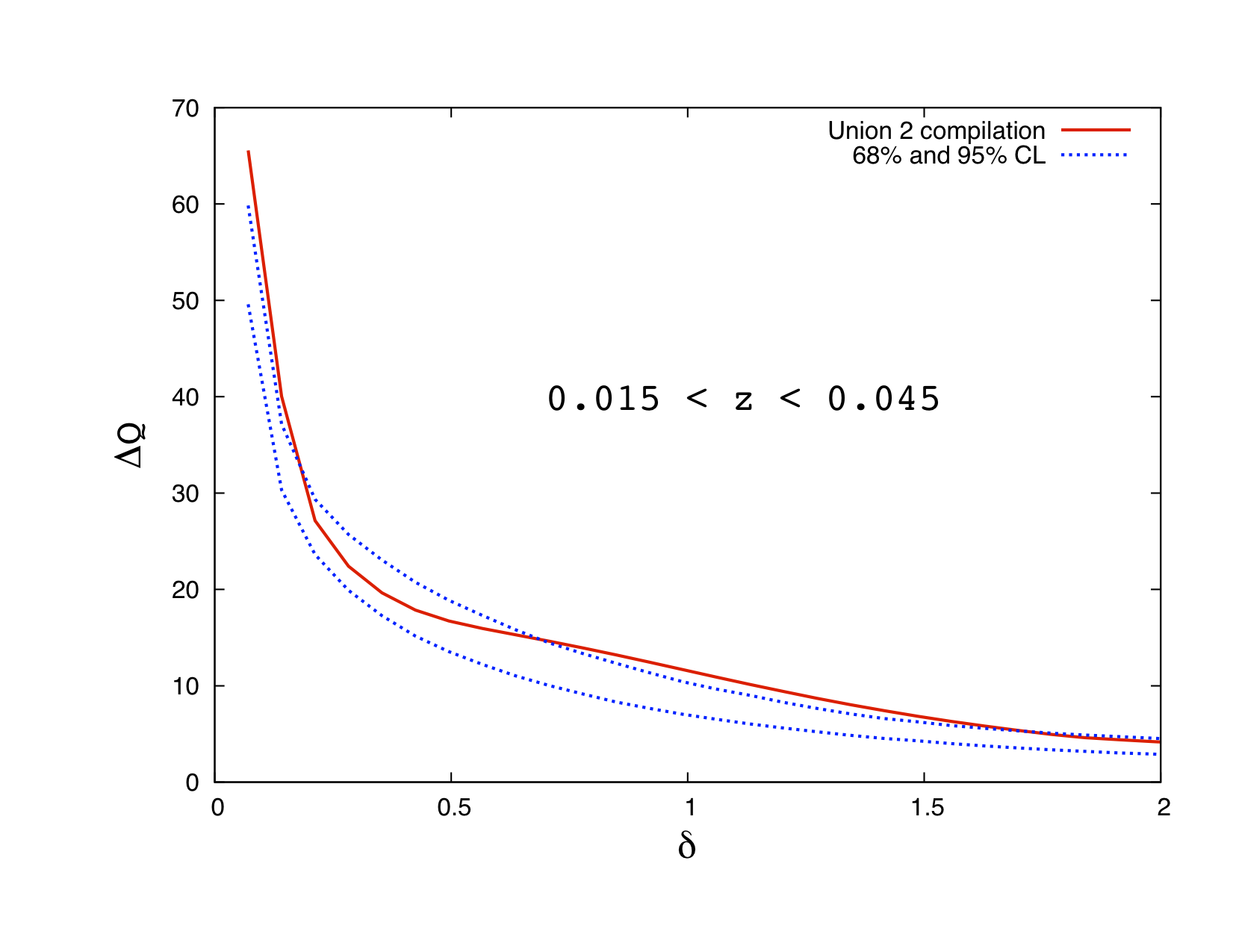}
\caption{The anisotropy measure
  $\Delta Q$ (eq.\ref{DeltaQ}) plotted against the smoothing length
  $\delta$, for different redshift shells. The full (red) lines show
  the results for the Union 2 SNe\,Ia data and the dotted (blue) lines
  are $68\%$ and $95\%$ confidence limits derived from 1000 Monte
  Carlo realisations. While each
  individual shell is consistent with isotropy, all shells have a
  mild anisotropy in the same direction so the cumulative result from
  combining them is inconsistent with isotropy at over
  $2\sigma$, as seen in the lower right panel.}
\label{fig:shells}
\end{figure*}

In Figure~\ref{fig:shells} we show results for the first 3 low redshift
shells and the cumulative result when data in all 3 shells is
combined. Although each individual shell is reasonably consistent with
isotropy, combining the data indicates an inconsistency of the
$\Lambda$CDM model with the data at $>2\sigma$. This happens because
the direction of the mild anisotropy in all 3 shells point the same
way. 

%---------------
%------------

We have added Table~\ref{table:results_dipole} to emphasise that the
important quantity at low-$z$ is not the P-value alone, since similar
P-values are obtained also at high redshift. One should consider also
the {\em consistency} of adjacent shells: the anisotropy in the first
3 shells point towards nearly the same direction. This strong
correlation makes the P-value for the cumulative result of the first 3
shells rather low: $z=0.045$. Figure~\ref {fig:shells} and
Table~\ref{table:results_dipole} show that at $z>0.1$ there are no
adjacent shells that point towards the same direction, hence, the
effect of these shells at high redshift cancel each other in the
cumulative result and the effect of first 3 shells dominates. The
importance of our differential shell analysis is now evident --- any
anisotropy in some specific redshift range would be smoothed out if
one were to use only the whole dataset.

%-------------
\begin{table*}
  \caption{Results from the residuals analysis for the P-value 
    which quantifies the level of agreement between the isotropic 
    universe and the data, for shells in redshift. On the
    left are the results for the individual shells and 
    on the right for the cumulative shells.} 
\begin{tabular}{c|c|c|c|c|c|c|c|c}
  \hline
  \hline
  $\Delta z$ & $b^\circ$ & $\ell^\circ$ & P-value & \vline & $\Delta z$ & $b^\circ$ & $\ell^\circ$ & P-value\\
  \hline
  0.015-0.025 & 46 & 265 & 0.140 & \vline & 0.015-0.025 & 46 & 265 & 0.140\\
  \hline
  0.025-0.035 & 2  & 320 & 0.112 & \vline & 0.015-0.035 & 27 & 297 & 0.039 \\
  \hline
  0.035-0.045 & 20 & 312 & 0.354 & \vline & 0.015-0.045 & 26 & 300 & 0.014\\
  \hline
  0.045-0.06  &-46 &  65 & 0.267 & \vline & 0.015-0.06  & 19 & 309 & 0.054\\
  \hline
  0.06-0.1   &-41 &  75 & 0.637 & \vline & 0.015-0.1   & 14 & 316 & 0.153\\
  \hline
  0.1-0.2     & -3 & 219 & 0.608 & \vline & 0.015-0.2   & 13 & 299 & 0.265\\
  \hline
  0.2-0.3     & 29 &  21 & 0.602 & \vline & 0.015-0.3   & 20 & 314 & 0.226\\
  \hline
  0.3-0.4     &-83 & 348 & 0.854 & \vline & 0.015-0.4   & 10 & 314 & 0.329\\
  \hline
  0.4-0.5     & 70 & 238 & 0.428 & \vline & 0.015-0.5   & 27 & 307 & 0.285 \\
  \hline
  0.5-0.6     & 16 &  15 & 0.177 & \vline & 0.015-0.6   & 27 & 326 & 0.180\\
  \hline
  0.6-0.8     &-77 &  45 & 0.108 & \vline & 0.015-0.8   &  3 & 332 & 0.279 \\
  \hline
  0.8-1.4     &-54 & 152 & 0.947 & \vline & 0.015-1.4   & -2 & 332 & 0.369\\
  \hline
\end{tabular}
\label{table:results_dipole}
\end{table*}
%---------------  

It is also possible to look for dipoles in different shells using our
method. We look for a direction in the sky for which the value of
$\Delta Q$ between that point $(\ell,b)$ and its opposite point on the
surface of the sphere --- i.e. $(\ell+180,-b)$ if $\ell<180$ and
$(\ell-180,-b)$ if $\ell>180$ --- is maximum. To calculate dipoles in
different shells we set $\delta=\pi/2$; for this value the weighted
average using gaussian smoothing is almost identical with the weighted
average by the inner product (i.e. $\cos\theta$) that is usually used to
find dipoles.

\begin{figure}
\includegraphics[width=0.5\textwidth]{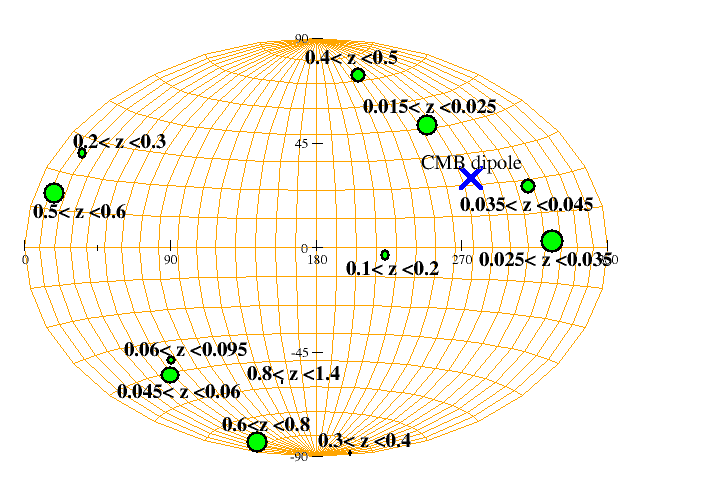}
\includegraphics[width=0.5\textwidth]{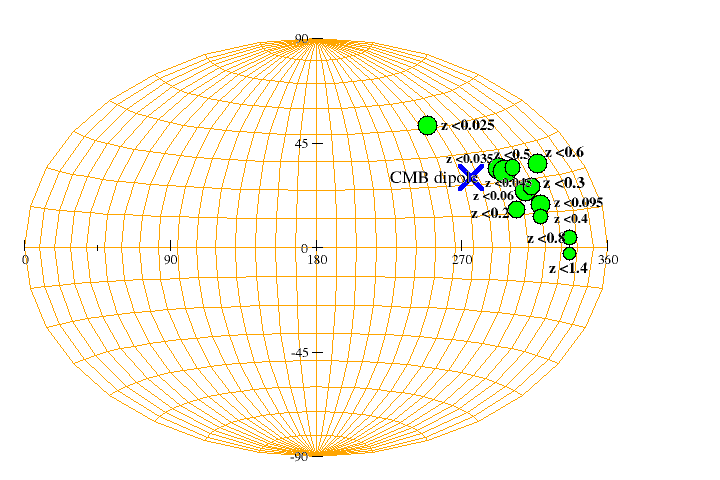}
\caption{Top panel: The dipole direction found by the residuals
  method in different redshift shells. Bottom panel:The cumulative
  dipole direction in shells of increasing radii. The size of the
  spots is proportional to the P-value; larger spots represent more
  significant disagreement between the data and the $\Lambda$CDM
  model.}
\label{fig:dipoles}
\end{figure}

In Figure~\ref{fig:dipoles} we see the positions of the dipoles in
galactic coordinates for different redshift shells. The size of the
spots are proportional to the P-values that are derived by comparing
the results from the real data to the 1000 Monte Carlo realizations of
the data when the assumed model is $\Lambda$CDM. Bigger spots indicate
a more anisotropic behavior or lower level of agreement between
$\Lambda$CDM and the real data (the size of the spots is proportional
to the P-value). In the bottom panel we see the cumulative results
which can help us to identify correlations between neighboring shells.

%----------------
%--------------

\begin{figure}
\center{
\includegraphics[width=0.49\textwidth]{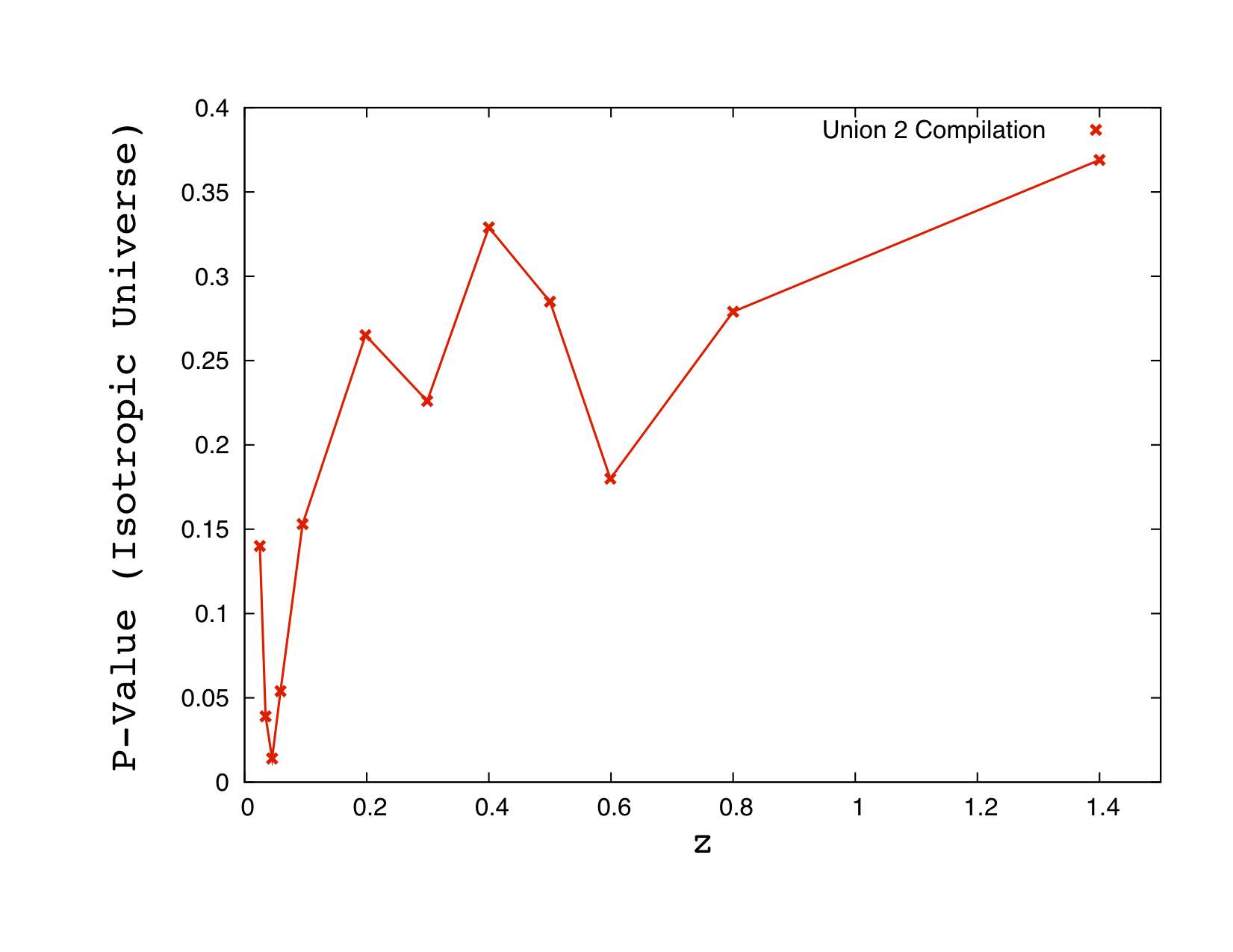}
\includegraphics[width=0.5\textwidth]{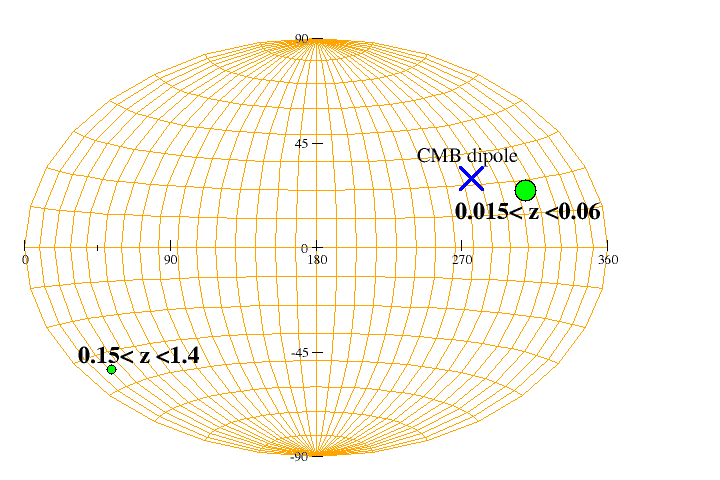}}
\caption{Top panel: P-value for the consistency of the isotropic
  universe with the data versus redshift. At $z \approx 0.05$ the
  P-value drops to 0.014 showing that isotropy is excluded at
  $3\sigma$ and we are not yet in the CMB rest frame.  Bottom panel:
  The cumulative analysis shows that at small redshift isotropy is
  excluded at 2--3 $\sigma$ with P = 0.054 for $0.015 < z < 0.06$;
  however at higher redshift agreement is achieved within 1 $\sigma$,
  with P = 0.594 for $0.15<z<1.4$.}
\label{fig:P-value}
\end{figure}

In Table~\ref{table:results_dipole} we see these results for different
shells, both individual and cumulative. It is interesting to see that
the P-value drops to 0.014 at $z<0.045$ i.e. a clear inconsistency
between $\Lambda$CDM (i.e. isotropy) and the data. The P-value
increases again with increasing redshift, suggesting that we are
approaching the CMB rest frame. This can be seen more clearly by
looking at the top panel of Figure~\ref{fig:P-value} where P-values
are plotted versus redshift for different cumulative shells. In the
bottom panel of Figure~\ref{fig:P-value} we can see the results from
our residuals analysis where we have divided the data into two
subsamples --- one for low redshifts $0.015 < z < 0.06$ and another
for higher redshifts $ 0.15 < z$. At small redshifts, the isotropic
universe lies over $\sigma$ away from the data with P=0.054 for $0.015
< z < 0.06$. However at higher redshift this discrepancy drops to
$\sim 1\sigma$ with P=0.594 for $z>0.15$.

%---------------
%-------------

\section{Test of isotropy and local bulk flow using Maximum Likelihood
  Analysis}
\label{sec:likelihood}

Now we use a different method to examine the low-redshift aniosotropy
detected above. We assume a bulk flow and use maximum likelihood
analysis method to find its value. We consider SNe\,Ia at $ z < 0.06$
(since the bulk flow would be negligible in relation to the Hubble
expansion rate at higher redshift) and minimise, for the Union 2 data:

\begin{eqnarray}
  \chi^2 (\Omega_{0m}, V_\mathrm{bulk}, j(\ell,b))= \\ \nonumber
  \sum_{i=1}^{N} \frac{\big[\mu_{i}^\mathrm{Union 2} - 5\log_{10} 
  \big(d_{\ell_i}^\mathrm{\Lambda CDM}(\Omega_\mathrm{0m}, H_0) 
  + \frac{V_\mathrm{bulk}.j (\ell, b)}{H_0} \big) - 25\big]^2}{\sigma_i^2}\,,
\end{eqnarray} 
where $N$ is the number of SNe\,Ia, $j(\ell, b)$ is a unit vector
representing the direction on the sky in Galactic coordinates,
$V_\mathrm{bulk}$ is the local bulk flow velocity in km\,s$^{-1}$,
$d_{\ell_i}^\mathrm{\Lambda CDM}(\Omega_\mathrm{0m}, H_0)$ is the
luminosity distance for the standard flat $\Lambda$CDM model and
$\Omega_\mathrm{0m}$ is the present day matter density (set equal to
0.27 although the choice is not important since the Hubble law is
linear here). It should be noted that in dealing with SNe\,Ia data,
$H_0$ is treated as a `nuisance parameter'. We consider different
cumulative redshift slices of the data, viz.  $0.015 < z < 0.025$,
$0.015 < z < 0.035$, $0.015 < z < 0.045$ and $0.015 < z < 0.06$.

\begin{figure}
\center{
\includegraphics[width=0.49\textwidth]{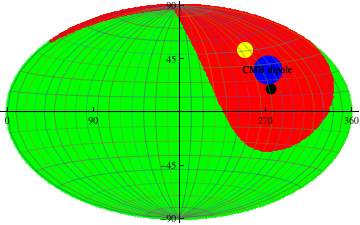}
\includegraphics[width=0.49\textwidth]{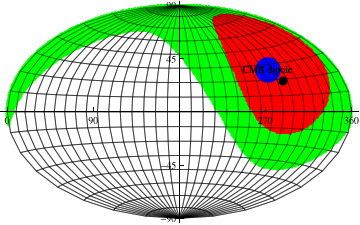}}
\caption{The top panel shows the dipole from the maximum likelihood
  analysis in the redshift band $0.015<z<0.025$ (61 SNe\,Ia). The best
  fit point is at ($b=16^\circ, \ell=271^\circ$) for $V_\mathrm{bulk}
  = 250$ km\,s$^{-1}$ and the red and green contours are the 1 and 2
  $\sigma$ confidence regions. The large blue spot is the direction of
  CMB dipole ($b=30^\circ\pm 2, \ell=276^\circ\pm 3$). The larger
  yellow spot, close to the CMB direction, is the best fit direction
  from the residuals analysis and the smaller black spot is the best
  fit direction from the maximum likelihood analysis (the magnitude of
  the dipole was given in Figure~\ref{fig:v-z}). The bottom panel
  shows the dipole for the redshift range $0.015<z<0.035$ (109
  SNe\,Ia). The blue spot is the CMB dipole and the black spot at
  $b=21^\circ, \ell=287^\circ$ is the best fit from the likelihood
  analysis for $V_\mathrm{bulk} = 260$ km\,s$^{-1}$, while the red and
  green patches are the 1 and 2 $\sigma$ confidence regions.}
\label{fig:ML2}
\end{figure}

\begin{figure}
\center{
\includegraphics[width=0.5\textwidth]{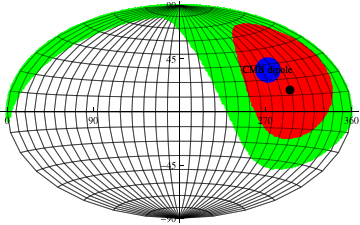}
\includegraphics[width=0.5\textwidth]{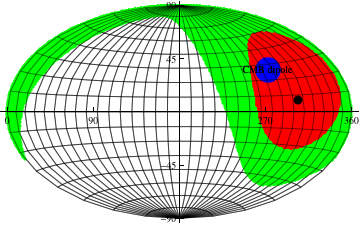}}
\caption{The top panel is for the range $0.015<z<0.045$ (127 SNe\,Ia)
  and the bottom panel is for $0.015<z<0.06$ (142 SNe\,Ia). The blue
  spot is the CMB dipole ($b=30^\circ\pm 2$, $\ell=276^\circ\pm 3$)
  and the black spots are the best-fit from the likelihood analysis at
  ($b=15^\circ, \ell=291^\circ$) for $V_\mathrm{bulk} = 270$
  km\,s$^{-1}$ for the top panel and at ($b=8^\circ, \ell=298^\circ$)
  for $V_\mathrm{bulk} = 260$ km\,s$^{-1}$ for the bottom panel, while
  the red and green patches are the 1 and 2 $\sigma$ confidence
  regions.}
\label{fig:ML3}
\end{figure}

In Figure~\ref{fig:ML2} we see that the data for $0.015 <z < 0.025$
(top panel) fits best with $V_\mathrm{bulk}= 260$ km\,s$^{-1}$ in the
direction $b=16^\circ$, $\ell=271^\circ$ which is close to the
direction of the CMB dipole ($b=30^\circ, \ell=276^\circ$)
\citep{Kogut:1993ag}. However, the $1\sigma$ contours are quite large
and an isotropic local universe with $V_\mathrm{bulk}=0$ is also
consistent with the data at $2\sigma$. When we assume a bigger
cumulative cut, the $1\sigma$ and $2\sigma$ regions shrink and we can
see that the isotropic universe is now shifted outside the $2\sigma$
region (Figure~\ref{fig:ML2} bottom panel). We can observe the same
trend up to $ 0.015 < z < 0.045$ where the 1 and 2 $\sigma$ regions
are still shrinking as seen in Figure~\ref{fig:ML3} (top
panel). However in the next step when we consider the data for $0.015
< z < 0.06$, the 1 and 2 $\sigma$ regions start becoming larger again
(Figure~\ref{fig:ML3}, bottom panel). These results are completely
consistent with our results from the smoothing and residuals
analysis. So we can conclude that the bulk of the local universe of
radius $z\sim 0.05$ is moving towards the CMB dipole direction with
$V_\mathrm{bulk}=260$ km\,s$^{-1}$. Our results are consistent with
\citet{Watkins:2008hf} who estimate a bulk flow of $V_\mathrm{bulk} =
416 \pm 78$ km\,s$^{-1}$ towards $b=60^\circ (+6, -6), \ell=282^\circ
(+11, -11)$ at a scale of $100 h^{-1}$~Mpc. On this scale we find
$V_\mathrm{bulk} = 250~(+190, -160)$ km\,s$^{-1}$ towards $b=21^\circ
(+34, -52), \ell=287^\circ (+62, -48)$, i.e. our results are
consistent within $1 \sigma$.

%---------------
%-------------

%---------------
%-------------

\begin{figure}
\includegraphics[width=0.5\textwidth]{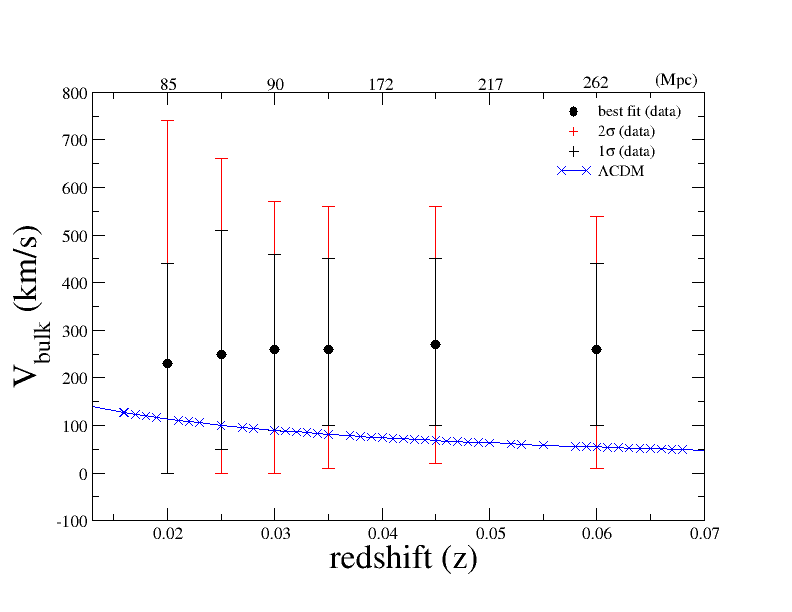}
\caption{The bulk flow as a function of redshift from the likelihood
 analysis. We see that a fast flow with $V_\mathrm{bulk} =
 260$ km\,s$^{-1}$ persists up to at least $z=0.06$ and systematically
 exceeds the peculiar velocity ($\sqrt 3$ times the 1-d RMS value)
 expected in $\Lambda$CDM normalised to WMAP-5 parameters
 \citep{Watkins:2008hf}. Note that the 1-d RMS expectation is shown
 (blue line) in order to emphasise that statistical analyses are carried
 out on the \emph{separate} components of the velocity, and also to
 facilitate comparison with the previous result of \citet{Watkins:2008hf}.}
\label{fig:v-z}
\end{figure}

In Figure~\ref{fig:v-z}, we see that the derived bulk flow using the
cumulative SNe\,Ia data is inconsistent with the prediction of a flat
$\Lambda$CDM model at 1--2 $\sigma$. This is a slight improvement
over using residuals (where the disagreement with $\Lambda$CDM was at
2--3 $\sigma$) the reason being that at small redshift one needs to
compare the data with the model at first-order in perturbation
theory. At high redshifts the comparison can be done directly between
the data and the unperturbed model since the perturbations have
subsided on large scales.

%---------------
%-------------

\section{Shapley infall}
\label{sec:shapley}

It is generally believed that our local bulk flow is due mainly to the
gravitational attraction of the Shapley supercluster which lies at
around $z \sim 0.04$ and that beyond Shapley the reference frame
should converge to the CMB rest frame.  We use the method of smoothing
and residuals, discussed in Section~\ref{sec:meth}, to study the
infall towards the Shapley concentration. Somewhat surprisingly the
SNe\,Ia data provides us with a clear picture of the infall.\

\begin{figure}
\includegraphics[width=0.5\textwidth]{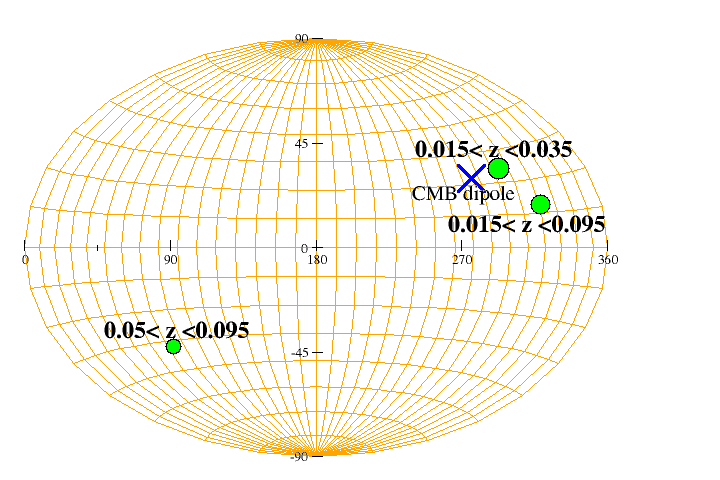}
\caption{ Shapley infall region as probed by the Union 2 SNe\,Ia
  data. The region in the redshift band $0.015<z<0.0345$ is falling
  towards Shapley in the direction shown by the big green spot
  (P=0.039 for the isotropic universe). However the shell
  $0.0522<z<0.095$ is falling towards Shapley in the opposite
  direction shown by another green spot (P=0.339 for the isotropic
  universe). Finally we show the cumulative result in the range
  $0.015<z<0.095$ --- the whole region is still moving in the
  direction of Shapley, as shown by the green spot ($b=14^\circ,
  \ell=316^\circ, P=0.153$). The CMB dipole direction is the large blue
  cross.}
\label{fig:shapley}
\end{figure}

We take Shapley to have an approximate extension of
$0.035<z<0.055$. First we consider all data in the redshift band
$0.015<z<0.0345$ which contains 109 SNe\,Ia and evaluate the dipole
which, as shown in the Figure~\ref{fig:shapley}, points towards
Shapley (which is approximately aligned with the direction of the CMB
dipole). While the movement towards Shapley seems to be strongly
favored by the data, the P-value for the isotropic universe is about
$P=0.039$. Next, we consider the data in the redshift range
$0.0522<z<0.095$ which contains 32 SNe\,Ia and as shown in
Figure~\ref{fig:shapley}, the direction now becomes completely {\em
  opposite}, indicating these SNe\,Ia are infalling towards
Shapley. The P-value for the isotropic universe in this range is
$P=0.339$.  Thus future precision SNe\,Ia data can provide us with
valuable information on the formation history of this giant
supercluster.

%%%%%%%
%%%%%%

\section{Conclusions}
\label{sec:conclusion} 

We have used SNe\,Ia data from the Union 2 catalogue to study the
(an)isotropy of our universe. Since the low and high redshift
anisotropy can in principle be of different origin, we have analysed
the data tomographically in different redshift slices, and also the
cumulative data set in order to trace any correlation between
different slices. We have developed a statistical tool of `residuals'
to examine the data in an unbiased manner at all redshifts. At low
redshift, we have performed a maximum likelihood analysis (which is
however unsuitable at high redshift). We find that an isotropic model
like $\Lambda$CDM is 2--3 $\sigma$ away from the data at
$z<0.06$. Although the agreement improves at high redshifts (to within
1 $\sigma$), we cannot single out $\Lambda$CDM as the preferred model
of the universe. The data becomes rather sparse at high redshift and
the error in distance measures increases, so the data may also agree
with alternative anisotropic models.

At low redshift, our results are rather robust and we find a bulk flow
of about 260 Km\,s$^{-1}$ in the direction of the Shapley
supercluster. We show that the Union 2 data provide the first evidence
of the infall on to Shapley; SNe\,Ia which are falling away from us
and towards Shapley are statistically dimmer that those which lie
beyond this supercluster and are falling towards us. We see no
indication of the decay of the bulk flow after Shapley which suggests
that the scale of anisotropy of our local universe is bigger than is
usually assumed and extends beyond $z\sim 0.1$.

Our analysis and results are important for the study of the expansion
history of the universe and the properties of dark energy. In all
SNe\,Ia compilations, an uncertainty of 300-500 Km\,s$^{-1}$ is
assumed for each data point to allow for bias introduced by random
peculiar velocities. However when there is a {\em coherent} motion of
SNe\,Ia towards a specific direction, this bias cannot be removed by
just increasing the size of the error bar (i.e. assuming the peculiar
velocities to be random). We will present in future work the effect of
this systematic motion of SNe\,Ia at low redshifts on the
reconstruction of the expansion history of the universe and estimation
of cosmological parameters like $q(z)$, $w(z)$ or
`Om'$(z)$~\citep{Sahni:2008,Shafieloo:2009ti}. 

We also note the
interesting observation by \citet{Tsagas:2009nh} that observers with
peculiar velocities have local expansion rates which are appreciably
different from the smooth Hubble flow, so can experience {\em
  apparently accelerated expansion} when the universe is actually
decelerating. Thus whether dark energy really needs to be invoked to
explain the SNe\,Ia Hubble diagram and other observations remains an
open question \citep{Sarkar:2007cx}.

\section*{Acknowledgments}

A.S., J.C. and R.M. thank the French ANR OTARIE for travel
support. A.S. and S.S. acknowledge the support of the EU FP6 Marie
Curie Research and Training Network ``UniverseNet''
(MRTN-CT-2006-035863) and the hospitality of the Institut
d'Astrophysique de Paris where part of this work was
undertaken. A.S. acknowledge the support of the Korea World Class
University grant no. R32-10130. We thank Mike Hudson, Robert Kirshner
and especially, Leandros Perivolaropoulos, for useful discussions and
acknowledge the use of the Union 2 data provided by M.~Blomqvist,
J.~Enander and E.~Mortsell ({\tt
  http://ttt.astro.su.se/~michaelb/SNdata/}).

\end{document}